\documentclass[onecolumn,preprintnumbers,amsmath,amssymb,eqsecnum,11pt]{revtex4}

\usepackage{epsf}
\usepackage{graphicx}  
\usepackage{dcolumn}   
\usepackage{bm}        

\begin{document}


\title{On the Entropy Function and the Attractor Mechanism for
Spherically Symmetric Extremal Black Holes}

 \author{Rong-Gen Cai\footnote{e-mail address:
cairg@itp.ac.cn}}
\address{
  Institute of Theoretical Physics, Chinese
Academy of Sciences, P.O. Box 2735, Beijing 100080, China}

\author{Li-Ming Cao\footnote{e-mail address:
caolm@itp.ac.cn}}
\address{Institute of Theoretical Physics, Chinese
Academy of Sciences,
 P.O. Box 2735, Beijing 100080, China\\
  Graduate School of the Chinese Academy of Sciences, Beijing 100039, China}

\begin{abstract}
In this paper we elaborate on the relation between the entropy
formula of Wald and the ``entropy function" method proposed by A.
Sen. For spherically symmetric extremal black holes, it is shown
that the expression of extremal black hole entropy given by A. Sen
can be derived from the general entropy definition of Wald, without
help of the treatment of rescaling the $AdS_2$ part of near horizon
geometry of extremal black holes. In our procedure, we only require
that the surface gravity approaches to zero, and it is easy to
understand the Legendre transformation of $f$, the integration of
Lagrangian density on the horizon, with respect to the electric
charges. Since the Noether charge form can be defined in an
``off-shell" form, we define a corresponding entropy function, with
which one can discuss the attractor mechanism for extremal black
holes with scalar fields.
\end{abstract}
\maketitle

\newpage
\section{Introduction}
The attractor mechanism  for extremal black holes has been studied
extensively in the past few years in supergravity theory and
superstring theory. It was initiated in the context supersymmetric
BPS black holes~\cite{FKS, AS, FK1, FK2, CY1, CY2} and generalized
to more general cases, such as supersymmetric black holes with
higher order corrections~\cite{CDM1, CDM2, CDM3, CDKM} and
non-supersymmetric attractors~\cite{FGK, GIJT, TT, GJMT, DST}.

Recently, A. Sen has proposed a so-called ``entropy function''
method for calculating the entropy of $n$-dimensional extremal black
holes, where the extremal black holes are defined to be the
space-times which have the near horizon geometry $AdS_{2}\times
S^{n-2}$ and corresponding isometry~\cite{Sen1, Sen2, Sen3, Sen4}.
It states that the entropy of such kind of extremal black holes can
be obtained by extremizing the ``entropy function'' with respect to
some moduli on the horizon, where the entropy function is defined as
$2\pi$ times the Legendre transformation ( with respect to the
electric charges ) of the integration of the Lagrangian over the
spherical coordinates on the horizon in the near horizon field
configurations. This method does not depend upon supersymmetry and
has been applied or generalized to many solutions in supergravity
theory, such as extremal black objects in higher dimensions,
rotating extremal black holes, various non-supersymmetric extremal
black objects and even near-extremal black holes~\cite{SSen1, SSen2,
SSen3, DSen, Pres, AE1, AE2, CPTY, Exi, Chand, SS1, SS2, AGJST,
CGLP, MS, AGM, CYY, CDM4, CP1, CP2, CP3, Garousi}.

In general, for spherically  symmetric extremal black holes in a
theory with Lagrangian $\mathcal{L}=\mathcal{L}(g_{ab}, R_{abcd},
\Phi_s, A_a^I)$,  the near horizon geometry of these black holes has
the form $AdS_2\times S^{n-2}$ ~\cite{Sen2, Sen3}. Due to
$SO(1,2)\times SO(n-1)$ isometry of this geometry, the field
configuration have the form as follows: The metric can be written
down as
\begin{equation}
\label{eq1}
ds^2=g_{ab}dx^adx^b=v_1\left(-\rho^2d\tau^2+\frac{1}{\rho^2}d\rho^2\right)+v_2
d\Omega_{n-2}^2\ ,
\end{equation}
where $v_1, v_2$ are constants which stand for the sizes of $AdS_2$
and $S^{n-2}$. Some other dynamical fields such as the scalar fields
and $U(1)$ gauge fields are also taken to be constant: $\Phi_s=u_s$
and $F_{\rho\tau}^I=e_I$. The magnetic-type fields are also fixed
with magnetic-charges $p_i$. Then, for this configuration, defining
\begin{equation}
\label{ffunction} f(v_1,v_2,u_s,e_I;p_i)=\int dx^2\wedge\cdots
\wedge dx^{n-1}\sqrt{-g}\mathcal{L}\, ,
\end{equation}
where the integration is taken on the horizon, and $\{x^2\cdots
x^{n-1}\}$ are angle coordinates of $S^{n-2}$, those constant moduli
can be fixed via the equations of motion
\begin{equation}
\frac{\partial f}{\partial v_1}=0,\quad \frac{\partial f}{\partial
v_2}=0, \quad \frac{\partial f}{\partial u_s}=0,\quad \frac{\partial
f}{\partial e_I}=q_I\, ,
\end{equation}
where $q_I$ are electrical-like charges for $U(1)$ gauge fields
$A_a^I$. To relate the entropy of the black holes to these
definitions, one defines $f_{\lambda}$ as (\ref{ffunction}) with the
Riemann tensor part in $\mathcal{L}$ multiplied by a factor
$\lambda$, and then one finds a relation between $f_{\lambda}$ and
the Wald formula for spherically symmetric black holes~\cite{wald}:
$S_{BH}=-2\pi
\partial f_{\lambda}/\partial \lambda|_{\lambda=1}$.
Consider the structure of the Lagrangian, one can find
\begin{equation}
\label{eqf} \lambda\frac{\partial f_{\lambda}}{\partial
\lambda}+v_1\frac{\partial f_{\lambda}}{\partial
v_1}+e_I\frac{\partial f_{\lambda}}{\partial e_I}-f_{\lambda}=0\, .
\end{equation}
When the equations of motion are satisfied, the entropy of black
holes turns out to be $ S_{BH}=2\pi(e_Iq_I-f) $.

 Therefore, one can
introduce the ``entropy function" for the extremal black holes
\begin{equation}
\label{entropyfunction}
\mathcal{E}(v_1,v_2,u_s,e_I;p_i)=2\pi\left(e_Iq_I-f(v_1,v_2,u_s,e_I;p_i)\right)\
,
\end{equation}
which is obtained by carrying an integral of the Lagrangian density
over $S^{n-2}$ and then taking the Legendre transformation with
respect to the electric fields $e_I$. For fixed electric changes
$q_I$ and magnetic charges $p_i$, these fields $u_s$ and $v_1$ and
$v_2$ are determined by extremizing the entropy function with
respect to the variables $u_s$ and $v_1$ and $v_2$. And then the
entropy of the extremal black holes is given by the extremum
 of the entropy function by substituting the values of $v_1$,
$v_2$ and $u_s$ back into the entropy function.  In addition, let us
notice that if the moduli fields $u_s$ are only dependent of the
charges $q_I$ and $p_i$, the attractor mechanism is then manifested,
and the entropy is a topological quantity.

This is a very simple and powerful method for calculating the
entropy of such kind of extremal black holes. In particular, one can
easily find the corrections to the entropy due to the higher
derivative terms in the effective action. However, we notice that
this method is established in a fixed coordinate system (\ref{eq1}).
If one uses another set of coordinates for the $AdS_2$ part, instead
of the coordinates $\{\rho,\tau \}$, it seems that one can not
define an entropy function as (\ref{entropyfunction}) because the
function $f$ is not invariant under the coordinate transformation.
In addition, the reason that to get the entropy of black holes, one
should do the Legendre transformation with respect to the electric
charges, but not include magnetic charges seems unclear in this
procedure. Some authors have pointed out that the entropy function
$\mathcal{E}$ resulting from this Legendre transformation of the
function $f$ with respect to electric charges transforms as a
function under the electric-magnetic dual, while the function $f$
does not~\cite{CDM4}. But it is not easy to understand the Legendre
transformation with respect to the angular-momentum $J$ in the
rotating attractor cases~\cite{AGJST}. There might be a more general
formalism for the entropy function, and the Legendre transformation
can be naturally understood in this frame. In this paper, we will
elaborate these issues in the ``entropy function" method and  show
that a general formalism of the ``entropy function" method  can be
extracted from the black hole entropy definition due to Wald {\it et
al.}~\cite{wald,Iwald,Iwald1}. In this procedure, we only require
that the surface gravity of the black hole approaches to zero. Our
entropy expression will reduce to the expression of A. Sen if we
choose the same coordinates as in~\cite{Sen2,Sen3}.

The extremal black holes are different objects from the non-extremal
ones due to different topological structures in Euclidean
sector~\cite{CTeitel, HHR, GK}. The extremal black hole has
vanishing surface gravity and has no bifurcation surface, so the
Noether charge method of Wald can not be directly used~\cite{wald}.
Thus, in this paper we regard the extremal black holes as the
extremal limit of non-extremal black holes as in~\cite{Sen2, Sen3,
SenKorea}. That is, we will first consider non-extremal black holes
and then take the extremal limit. In this sense, the definitions of
Wald are applicable.

The paper is organized as follows. In section II, we make a brief
review on the entropy definition of Wald and give the required
formulas. In section III, we give the near horizon analysis for the
extremal black holes and derive the general form of the entropy. In
section IV, we define the entropy function and discuss the attractor
mechanism for the black holes with various moduli fields. The
conclusion and discussion are given in section V.

\section{The Definition of Wald}

In differential covariant theories of gravity, Wald showed that the
entropy of a black hole is a kind of Noether charge \cite{wald,
Iwald}. In this paper, we will use the Wald's method to define the
entropy functions for spherically symmetric black holes. Assume the
differential covariant Lagrangian of $n$-dimensional space-times
$(M,g_{ab})$ is
\begin{equation}
\label{Lagrangian}
\mathbf{L}=\mathcal{L}(g_{ab}, R_{abcd},
\Phi_{s}, A_{a}^{I}) ~\mbox{{\boldmath $\epsilon$}},
\end{equation}
where we have put the Lagrangian in the form of differential form
and $\mbox{{\boldmath $\epsilon$}}$ is the volume element.
$R_{abcd}$ is Riemann tensor (since we are mainly concerning with
extremal black holes, therefore we need not consider the covariant
derivative of the Riemann tensor). $\{\Phi_s$, $s=0,1,\cdots\}$ are
scalar fields,  $\{A_a^I$, $I=1,\cdots\}$ are $U(1)$ gauge
potentials, and the corresponding gauge fields are
$F^I_{ab}=\partial_a A_b^I-\partial_b A_a^I$. We will not consider
the Chern-Simons term as~\cite{Sen3}.

The variation of the Lagrange density $\mathbf{L}$ can be written as
\begin{equation}
\delta\mathbf{L}=\mathbf{E}_{\psi}\delta \psi +d \mathbf{\Theta},
\end{equation}
where $\mathbf{\Theta}=\mathbf{\Theta}(\psi,\delta \psi)$ is an
$(n-1)$-form, which is called {\it symplectic potential form}, and
it is a local linear function of field variation (we have denoted
the dynamical fields as $\psi=\{g_{ab},\Phi_{s}, A_{a}^{I}\}$).
$\mathbf{E}_{\psi}$ corresponds to the equations of motion for the
metric and other fields. Let $\xi$ be any smooth vector field on the
space-time manifold, then one can define a {\it Noether current
form} as
\begin{equation}
\label{Noethercurrent}
\mathbf{J}[\xi]=\mathbf{\Theta}(\psi,\mathcal{L}_{\xi} \psi)-\xi
\cdot \mathbf{L}\, ,
\end{equation}
where $``\cdot"$ means the inner product of a vector field with a
differential form, while $\mathcal{L}_{\xi}$ denotes the Lie
derivative for the dynamical fields. A standard calculation gives
\begin{equation}
d\mathbf{J}[\xi]=-\mathbf{E}\mathcal{L}_{\xi}\psi\, .
\end{equation}
It implies that $\mathbf{J}[\xi]$ is closed when the equations of
motion are satisfied. This indicates that there is a locally
constructed $(n-2)$-form $\mathbf{Q}[\xi]$ such that, whenever
$\psi$ satisfy the equations of motion, we have
\begin{equation}
\mathbf{J}[\xi]=d\mathbf{Q}[\xi]\, .
\end{equation}
In fact, the {\it Noether charge form} $\mathbf{Q}[\xi]$ can be
defined in the so-called ``off shell" form so that the Noether
current $(n-1)$-form can be written as~\cite{Iwald1}
\begin{equation}
\mathbf{J}[\xi]=d\mathbf{Q}[\xi]+\xi^a\mathbf{C}_a\, ,
\end{equation}
where $\mathbf{C}_a$ is locally constructed out of the dynamical
fields in a covariant manner. When the equations of motion hold,
$\mathbf{C}_a$ vanishes.  For general stationary black holes, Wald
 has shown that the entropy of the black holes is a  Noether
charge~\cite{wald}, and may be expressed as
\begin{equation}
S_{BH}=2\pi \int _{\mathcal{H}} \mathbf{Q}[\xi]\, ,
\end{equation}
here $\xi$ be the Killing field which vanishes on the bifurcation
surface of the black hole. It should be noted that  the Killing
vector field has been normalized here so that the surface gravity
equals to ``1". Furthermore, it was shown in~\cite{Iwald} that the
entropy can also be put into a form
\begin{equation}
\label{waldentropy}
S_{BH}=-2\pi\int_{\mathcal{H}}E_R^{abcd}\mbox{{\boldmath
$\epsilon$}}_{ab}\mbox{{\boldmath $\epsilon$}}_{cd},
\end{equation}
where $\mbox{{\boldmath $\epsilon$}}_{ab}$ is the binormal to the
bifurcation surface $\mathcal{H}$, while $E_R^{abcd}$ is the
functional derivative of the Lagrangian with respect to the Riemann
tensor with metric held fixed. This formula is purely geometric and
does not include the surface gravity term.  In this paper, since we
will treat a limit procedure with surface gravity approaching to
zero, we will not normalize the Killing vector such that the surface
gravity equal to one. So we use the formula (\ref{waldentropy}) to
define the entropy of black holes as in~\cite{Sen2, Sen3, SenKorea}.
For an asymptotically flat, static spherically symmetric black hole,
one can simply choose $\xi=\partial_t=\frac{\partial}{
\partial t}$.

For the Lagrangian as (\ref{Lagrangian}), we have
\begin{equation}
\delta \mathbf{L}=\mathbf{E}^{ab}\delta
g_{ab}+\mathbf{E}^{a}_{I}\delta A_{a}^{I}+\mathbf{E}^{s}\delta
\Phi_{s}+d\mathbf{\Theta}\, ,
\end{equation}
where
\begin{equation}
\mathbf{E}^{a}_{I}=-2\mbox{{\boldmath
$\epsilon$}}\nabla_{b}\left(\frac{\partial \mathcal{L}}{\partial
F_{ab}^{I}}\right)\, ,
\end{equation}
\begin{equation}
\mathbf{E}^{s}=\mbox{{\boldmath
$\epsilon$}}\left(\frac{\partial\mathcal{L}}{\partial
\Phi_{s}}-\nabla_{a}\frac{\partial\mathcal{L}}{\partial
\nabla_{a}\Phi_s}\right)\, ,
\end{equation}
\begin{equation}
\mathbf{E}^{ab}=\mbox{{\boldmath $\epsilon$}}\left(\frac{\partial
\mathcal{L}}{\partial
g_{ab}}+\frac{1}{2}g^{ab}\mathcal{L}+\frac{\partial
\mathcal{L}}{\partial
R_{cdea}}R_{cde}{}{}^b+2\nabla_{c}\nabla_{d}\frac{\partial
\mathcal{L}}{\partial R_{cabd}}\right)
\end{equation}
are the equations of motion for the $U(1)$ gauge fields,  the scalar
fields and the metric $g_{ab}$, respectively.  The symplectic
potential form has the form
\begin{eqnarray}
\mathbf{\Theta}_{a_1\cdots a_{n-1}}&=&\Bigg{(}\frac{\partial
{\mathcal{L}}}{\partial\nabla_{a} \Phi_s}\delta
\Phi_s+2\frac{\partial \mathcal{L}}{\partial F^{I}_{ab}}\delta
A_{b}^{I}\nonumber \\
&+&2\frac{\partial\mathcal{L}}{\partial R_{abcd}}\nabla_d \delta
g_{bc}-2\nabla_{d}\frac{\partial\mathcal{L}}{\partial
R_{dbca}}\delta g_{bc}\Bigg{)}\mbox{{\boldmath
$\epsilon$}}_{aa_1\cdots a_{n-1}}\, .
\end{eqnarray}
Let $\xi$ be an arbitrary vector field on the space-time, The Lie
derivative of $\xi$ on the fields are
\begin{equation}
\mathcal{L}_{\xi}\Phi_s=\xi^a\nabla_{a}\Phi_s,\quad
\mathcal{L}_{\xi}g_{ab}=\nabla_{a}\xi_b+\nabla_b\xi_a\,
,\quad\mathcal{L}_{\xi}A_{a}^I=\nabla_{a}(\xi^b
A_b^{I})+\xi^bF^{I}_{ba}\, .
\end{equation}
Substituting these Lie derivatives into the symplectic potential
form, we find
\begin{eqnarray}
\mathbf{\Theta}_{a_1\cdots a_{n-1}}&=&\Bigg{[}\frac{\partial
{\mathcal{L}}}{\partial\nabla_{a}
\Phi_s}\xi^b\nabla_{b}\Phi_s+2\frac{\partial \mathcal{L}}{\partial
F^{I}_{ab}}\nabla_{b}(\xi^c A_c^{I})+2\frac{\partial
\mathcal{L}}{\partial F^{I}_{ab}}\xi^cF^{I}_{cb}\nonumber
\\
&+&2\frac{\partial\mathcal{L}}{\partial R_{abcd}}\nabla_d
(\nabla_b\xi_c+\nabla_c\xi_b)-2\nabla_{d}\frac{\partial\mathcal{L}}{\partial
R_{dbca}}(\nabla_b\xi_c+\nabla_c\xi_b)\Bigg{]}\mbox{{\boldmath
$\epsilon$}}_{aa_1\cdots a_{n-1}}\nonumber\\
&=&\Bigg{[}\frac{\partial {\mathcal{L}}}{\partial\nabla_{a}
\Phi_s}\xi^b\nabla_{b}\Phi_s+2\nabla_{b}\left(\frac{\partial
\mathcal{L}}{\partial F^{I}_{ab}}\xi^c
A_c^{I}\right)-2\nabla_{b}\frac{\partial \mathcal{L}}{\partial
F^{I}_{ab}}\xi^c A_c^{I}+2\frac{\partial \mathcal{L}}{\partial
F^{I}_{ab}}\xi^cF^{I}_{cb}\nonumber
\\
&+&2\frac{\partial\mathcal{L}}{\partial R_{abcd}}\nabla_d
(\nabla_b\xi_c+\nabla_c\xi_b)-2\nabla_{d}\frac{\partial\mathcal{L}}{\partial
R_{dbca}}(\nabla_b\xi_c+\nabla_c\xi_b)\Bigg{]}\mbox{{\boldmath
$\epsilon$}}_{aa_1\cdots a_{n-1}}\, .
\end{eqnarray}
Then, we have
\begin{eqnarray}
\mathbf{\Theta}_{a_1\cdots
a_{n-1}}&=&\Bigg{[}2\nabla_{b}\left(\frac{\partial
\mathcal{L}}{\partial F^{I}_{ab}}\xi^c
A_c^{I}\right)-\nabla_b\left(\frac{\partial\mathcal{L}}{\partial
R_{abcd}}\nabla_{[c}\xi_{d]}\right)\Bigg{]}\mbox{{\boldmath
$\epsilon$}}_{aa_1\cdots a_{n-1}}+\cdots \nonumber \\
&+&\Bigg{[}\frac{\partial {\mathcal{L}}}{\partial\nabla_{a}
\Phi_s}\xi^b\nabla_{b}\Phi_s+2\frac{\partial \mathcal{L}}{\partial
F^{I}_{ab}}\xi^cF^{I}_{cb}+\cdots\cdots\Bigg{]}\mbox{{\boldmath
$\epsilon$}}_{aa_1\cdots a_{n-1}}\nonumber
\\
&-&2\nabla_{b}\frac{\partial \mathcal{L}}{\partial F^{I}_{ab}}\xi^c
A_c^{I}\mbox{{\boldmath $\epsilon$}}_{aa_1\cdots a_{n-1}}\, .
\end{eqnarray}
The first line in the above equation will give the Noether charge
form, while the second line together with  the terms in $\xi\cdot
\mathbf{L}$ in Eq. (\ref{Noethercurrent}) will give the constraint
which corresponds to the equations of motion for the metric. For
example, the first term in the second line combined with scalar
fields terms in $\xi\cdot \mathbf{L}$ will give the energy-momentum
tensor for scalar fields. Similarly the second term in the second
line will enter the energy-momentum tensor for the $U(1)$ gauge
fields in the equations of motion for the metric. The last line in
the above equation will give the constraint which corresponds to the
equations of motion for the $U(1)$ gauge fields. Thus, we find
\begin{equation}
\mathbf{J}[\xi]=d\mathbf{Q}[\xi]+ \xi^a\mathbf{C}_a\, ,
\end{equation}
where
\begin{equation}
\label{conservedcharge}
\mathbf{Q}=\mathbf{Q}^{F}+\mathbf{Q}^g+\cdots
\end{equation}
with
\begin{equation}
\mathbf{Q}^F_{a_1\cdots a_{n-2}}=\frac{\partial
\mathcal{L}}{\partial F^{I}_{ab}}\xi^c A_c^{I}\mbox{{\boldmath
$\epsilon$}}_{aba_1\cdots a_{n-2}}\, ,
\end{equation}
\begin{equation}
\mathbf{Q}^g_{a_1\cdots
a_{n-2}}=-\frac{\partial\mathcal{L}}{\partial
R_{abcd}}\nabla_{[c}\xi_{d]}\mbox{{\boldmath
$\epsilon$}}_{aba_1\cdots a_{n-2}}\, .
\end{equation}
The $``\cdots"$ terms are not important for our following
discussion, so we brutally drop them at first. We will give a
discussion at the end of the next section for these additional
terms. Especially, the constraint for the $U(1)$ gauge fields is
simply
\begin{equation}
\label{U1constrain} \mathbf{C}^F_{ca_1\cdots
a_{n-1}}=-2\nabla_{b}\frac{\partial \mathcal{L}}{\partial
F^{I}_{ab}}A_c^{I}\mbox{{\boldmath $\epsilon$}}_{aa_1\cdots
a_{n-1}}\, .
\end{equation}
The  term $\mathbf{Q}^F$ in the $\mathbf{Q}$ was not discussed
explicitly in the earlier works of Wald {\it et al.}~\cite{wald,
Iwald, Iwald1}. This is because that the killing vector vanishes on
the bifurcation surface and the dynamical fields are assumed to be
smooth on the bifurcation surface. However, in general,  the $U(1)$
gauge fields are singular on the bifurcation surface, so one have to
do a gauge transformation, $A\rightarrow A'= A-A|_{\mathcal{H}}$,
such that the $\xi^a A'_a$ are vanished on this surface, and then
$\mathbf{Q}^F$. This gauge transformation will modify the data of
gauge potential at infinity and an additional potential-charge term
$\Phi \delta Q$ into the dynamics of the charged black holes from
infinity, where $\Phi=\xi^cA_c|_{\mathcal{H}}$ is the  electrostatic
potential on the horizon of the charged black hole and $Q$ is the
electric charge~\cite{Swald}. Another treatment is: We only require
the smoothness of the gauge potential projecting on the bifurcation
surface, i.e., $\xi^aA_a$ instead of the gauge potential itself, so
$\mathbf{Q}^F$ will generally not vanish on the bifurcation surface,
and then $\Phi=\xi^cA_c|_{\mathcal{H}}$ is introduced into the law
of black hole without help of gauge transformation~\cite{SGao}.
Similarly, in the next sections of this paper we only require that
the projection of the gauge potential on the bifurcation surface is
smooth. Since our final result will not depend on the gauge
potential, the gauge transformation mentioned above will not effect
our discussion. One can do such gauge transformation if necessary.
In this paper, however, we will merely use the explicit form of the
Noether charge $(n-2)$-form and we will not discuss the first law.
Certainly, it is interesting to give a general discussion on the
thermodynamics of these black holes. The relevant discussion can be
found in a recent paper~\cite{SurWalper}.

\section{Entropy of extremal black holes}
In this section, we will use the formulas above to give the general
entropy function for static spherically symmetric extremal black
holes. Assume that the metric for these black holes is of the form
\begin{equation}
\label{generalmetric}
ds^2=-N(r)dt^2+\frac{1}{N(r)}dr^2+\gamma(r)d\Omega_{n-2}^2\, ,
\end{equation}
where $N, \gamma$ are functions of radial coordinate $r$, and
$d\Omega_{n-2}^2$ is the line element for the $(n-2)$-dimensional
sphere. The horizon $r=r_H$ corresponds to $N(r_H)=0$. If the
equations of motion are satisfied, the constraint $\mathbf{C}_a=0$,
and we have
$$
\mathbf{J}[\xi]=d\mathbf{Q}[\xi]\, .
$$
Consider a near horizon region ranged from $r_H$ to $r_H+\Delta r$,
we have
\begin{eqnarray}
\label{nhint} &&\int_{r_{H}+\Delta
r}\mathbf{Q}[\xi]-\int_{r_{H}}\mathbf{Q}[\xi]=\int_{\mathcal{H}\times
\Delta r}\mathbf{J}[\xi]\nonumber \\
&&=\int_{\mathcal{H}\times \Delta r}\mathbf{\Theta}-\xi\cdot
\mathbf{L}\, .
\end{eqnarray}
If $\xi$ is a Killing vector, then $\mathbf{\Theta}=0$, and
\begin{equation}
\int_{r_{H}+\Delta
r}\mathbf{Q}[\xi]-\int_{r_{H}}\mathbf{Q}[\xi]=-\int_{\mathcal{H}\times
\Delta r}\xi\cdot \mathbf{L}\, .
\end{equation}
Thus we arrive at
\begin{eqnarray}
\label{intnearhorzion} &&\int_{r_{H}+\Delta
r}\mathbf{Q}^g[\xi]-\int_{r_{H}}\mathbf{Q}^g[\xi]\nonumber
\\
&=&-\int_{r_{H}+\Delta
r}\mathbf{Q}^F[\xi]+\int_{r_{H}}\mathbf{Q}^F[\xi]-\int_{\mathcal{H}\times
\Delta r}\xi\cdot \mathbf{L}\, .
\end{eqnarray}
Taking $\xi=\partial_t$, (since we consider the asymptotically flat
space-time, $N(r)$ has the property $\lim_{r\rightarrow \infty}
N(r)=1$, such that $\partial_t$ has a unit norm at infinity.), we
have $\nabla_{[a}\xi_{b]}=\frac{1}{2}N'\mbox{{\boldmath
$\epsilon$}}_{ab}$, and
\begin{eqnarray}
\label{equation1} &&\int_{r_{H}+\Delta
r}\mathbf{Q}^g[\partial_t]-\int_{r_{H}}\mathbf{Q}^g[\partial_t]\nonumber\\
&=&\frac{1}{2}\left[N'(r_H+\Delta
r)B(r_H+\Delta r)-N'(r_H)B(r_H)\right]\nonumber \\
&=&\frac{1}{2}\Delta
r\left[N''(r_H)B(r_H)+N'(r_H)B'(r_H)\right]+\mathcal{O}(\Delta
r^2)\, ,
\end{eqnarray}
where we have defined a function $B(r)$
\begin{equation}
B(r) \equiv
-\int_{r}\frac{1}{(n-2)!}\frac{\partial\mathcal{L}}{\partial
R_{abcd}}\mbox{{\boldmath $\epsilon$}}_{cd}\mbox{{\boldmath
$\epsilon$}}_{aba_1\cdots a_{n-2}}dx^{a_1}\wedge\cdots\wedge
dx^{a_{n-2}}\, .
\end{equation}
Note that the $\mathbf{Q}^F$ terms in the right hand side of Eq.
(\ref{intnearhorzion}) can be written as
\begin{eqnarray}
\label{equation2} &&-\int_{r_{H}+\Delta
r}\mathbf{Q}^F[\partial_t]+\int_{r_{H}}\mathbf{Q}^F[\partial_t]\nonumber
\\
&=&A_t^I(r_H+\Delta r)q_I -A_t^I(r_H)q_I\nonumber \\
&=& q_I {A'}_t^I(r_H)\Delta r +{\cal O}(\triangle r^2)\nonumber \\
&=& q_I F_{rt}^I(r_H)\Delta r+{\cal O}(\triangle r^2)=q_I
\tilde{e}_I\Delta r+{\cal O}(\triangle r^2)\, ,
\end{eqnarray}
where $A^I_t=(\partial_t)^a A^I_a$, $\tilde{e}_I \equiv
F_{rt}^I(r_H)$, and the $U(1)$ electrical-like charges are defined
to be
\begin{equation}
\label{eq34}
 q_I=-\int_{r}\frac{1}{(n-2)!}\frac{\partial
\mathcal{L}}{\partial F^{I}_{ab}}\mbox{{\boldmath
$\epsilon$}}_{aba_1\cdots a_{n-2}}dx^{a_1}\wedge\cdots \wedge
dx^{a_{n-2}}\, .
\end{equation}
They do not change with the radii $r$. This is ensured by the
Gaussian law.   Note that there is an integration on the sphere part
in (\ref{eq34}), therefore the only $F^I_{rt}$ in $F^I_{ab}$ is
relevant, so that we can simply write $F^{I}_{ab}(r_H)$ as
$-\tilde{e}_I\mbox{{\boldmath $\epsilon$}}_{ab}$. Considering
$-2\tilde{e_I}^2=\tilde{e}_I\mbox{{\boldmath
$\epsilon$}}_{ab}\tilde{e}_I\mbox{{\boldmath $\epsilon$}}^{ab}$ we
have
\begin{equation}
\frac{\partial \mathcal{L}}{\partial
F^{I}_{ab}}=-\frac{\partial\mathcal{L}}{\partial
\tilde{e}_I}\frac{\partial
\tilde{e}_I}{\partial(\tilde{e}_I\mbox{{\boldmath
$\epsilon$}}_{ab})}=\frac{1}{2}\frac{\partial\mathcal{L}}{\partial
\tilde{e}_I}\mbox{{\boldmath $\epsilon$}}^{ab}\, .
\end{equation}
Substituting this result into the definition of the electric
charges, we find
\begin{equation}
q_I=-\frac{\partial }{\partial
\tilde{e}_I}\int_{r_H}\frac{\mathcal{L}}{2(n-2)!}\mbox{{\boldmath
$\epsilon$}}^{ab}\mbox{{\boldmath $\epsilon$}}_{aba_1\cdots
a_{n-2}}dx^{a_1}\wedge\cdots \wedge
dx^{a_{n-2}}=\frac{\partial\tilde{f}(r_H)}{\partial\tilde{e}_I}\, .
\end{equation}
Here $\tilde{f}(r_H)$ will be defined below in Eq.
(\ref{ffunction1}). The last term in the right hand side of  Eq.
(\ref{intnearhorzion}) can be written as
\begin{equation}
\int_{\mathcal{H}\times \Delta r}\partial_t\cdot
\mathbf{L}=\int_{r_H}^{r_H+\Delta r}dr\int dx^2\wedge\cdots \wedge
dx^{n-1} \sqrt{-g}\mathcal{L}=\int_{r_H}^{r_H+\Delta r}dr
\tilde{f}(r)\, ,
\end{equation}
where
\begin{equation}
\label{ffunction1} \tilde{f}(r)=\int_{r} dx^2\wedge\cdots\wedge
dx^{n-1}\sqrt{-g}\mathcal{L}\, .
\end{equation}
Thus we arrive at
\begin{equation}
\label{equation3} \int_{\mathcal{H}\times \Delta r}\partial_t\cdot
\mathbf{L} = \Delta r \tilde{f}(r_H)+{\cal O}(\triangle r^2)\, ,
\end{equation}
up to the leading order of $\triangle r$.   Substituting Eqs.
(\ref{equation1}), (\ref{equation2}) and (\ref{equation3}) into Eq.
(\ref{intnearhorzion}), we get
\begin{eqnarray}
 &&\frac{1}{2}\Delta
r\left[N''(r_H)B(r_H)+N'(r_H)B'(r_H)\right]+\mathcal{O}(\Delta
r^2)\nonumber\\
&=&\Delta r q_I \tilde{e}_I-\Delta r \tilde{f}(r_H)\, .
\end{eqnarray}
Considering the limit  $\Delta r\rightarrow 0$, we find
\begin{eqnarray}
\label{nonextremal}
\frac{1}{2}\left[N''(r_H)B(r_H)+N'(r_H)B'(r_H)\right]=
 q_I \tilde{e}_I-\tilde{f}(r_H)\, .
\end{eqnarray}
So far, we have not specialized to extremal black holes; therefore,
the above results hold for general non-extremal black holes. For the
extremal black holes limit with $N'(r_H) \rightarrow 0$, while
$N''(r_H)\neq 0$, from (\ref{nonextremal}) we have
\begin{equation}
B(r_H)= \frac{2}{N''(r_H)}\left(q_I
\tilde{e}_I-\tilde{f}(r_H)\right)\, .
\end{equation}
Since we view the extremal black holes as the extremal limit of
non-extremal black holes, the entropy formula of Wald is applicable
for the extremal black holes. Note that $B(r_H)$ is nothing but the
integration in Eq. (\ref{waldentropy}) without the $2\pi$ factor.
Thus, the entropy of the extremal black holes can be expressed as
\begin{equation}
\label{entropynew} S_{BH}=2\pi B(r_H)=
\frac{4\pi}{N''(r_H)}\left(q_I \tilde{e}_I-\tilde{f}(r_H)\right)\, .
\end{equation}
This is one of main results in this paper.  It is easy to see that
this entropy form is very similar to the one in the ``entropy
function" method of A. Sen. But some remarks are in order:

\indent(i). We have not stressed that  the extremal black holes have
the near horizon geometry $AdS_2\times S^{n-2}$ as in~\cite{Sen2,
Sen3} although the vanishing surface gravity and the metric
assumption~(\ref{generalmetric}) may coincide with the definition
through the near horizon geometry.  However, let us notice that some
extremal black holes have near horizon geometries of the form
$AdS_3$ products some compact manifold $X$. In our procedure, the
near horizon geometry is not necessary to be $AdS_2\times S^{n-2}$
and the only requirement is to have vanishing surface gravity.
Therefore our procedure can be used to discuss that kind of extremal
black holes whose near horizon geometry is of the form $AdS_3\times
X$ by simply modifying  the metric assumption
in~Eq.(\ref{generalmetric}).

\indent(ii). Our result is explicitly invariant under coordinate
transformation, and this can be easily seen from the above process.
We have not used the treatment method Eq.(\ref{eqf}) employed by A.
Sen.

\indent(iii). The Legendre transformation with respect to the
electric charges appears naturally in this procedure, while the
Legendre transformation with respect to the magnetic charges does
not appear.

\indent(iv). If we choose a set of coordinates as the one
in~\cite{Sen2,Sen3}, our expression for the entropy is exactly same
as the one given by A. Sen. This can be seen as follows. In the
extremal limit $N'(r_H)= 0$, we can rewrite the metric near the
horizon as
\begin{equation}
ds^2=-\frac{1}{2}N''(r_H)(r-r_H)^2dt^2+\frac{2}{N''(r_H)(r-r_H)^2}dr^2+\gamma(r_H)d\Omega_{n-2}^2\,
.
\end{equation}
Redefine the coordinates as
\begin{equation}
\rho=r-r_H,\quad \tau=\frac{1}{2}N''(r_H) t\, .
\end{equation}
Then, the near horizon metric can be further rewritten as
\begin{equation}
\label{nearhorizongeometry1}
ds^2=\frac{2}{N''(r_H)}\left(-\rho^2d\tau^2+\frac{1}{\rho^2}d\rho^2\right)+\gamma(r_H)d\Omega_{n-2}^2\,
.
\end{equation}
The components of gauge fields $F^I_{rt}$ and $\tilde{f}$ are
dependent of coordinates, in this new set of  coordinates they are
\begin{equation}
\tilde{e}_I=\frac{1}{2}N''(r_H) e_I\, ,
\end{equation}
\begin{equation}
\tilde{f}(r_H)=\frac{1}{2}N''(r_H) f\, .
\end{equation}
where
\begin{equation}
e_I=F^I_{\rho \tau}(r_H),\quad f=\int_{r_H} dx^2\wedge\cdots\wedge
dx^{n-1}\sqrt{-g'}\mathcal{L}\, .
\end{equation}
Since the entropy is invariant under the coordinate transformation,
we find in these coordinates like $\{\tau,\rho,\cdots \}$,
\begin{equation}
\label{entropyfunction1} S_{BH}=2\pi\left(q_I e_I-f\right)\, .
\end{equation}
This is nothing but the entropy formula given by A. Sen for extremal
black holes. Since the factor $2/N''(r_H)$
 in (\ref{entropynew}) disappears in this new set of  coordinates, the entropy
formula becomes more simple and good look. This is an advantage of
this set of coordinates. But we  would like to stress that the
entropy expression with the factor $``2/N''(r_H)"$ makes it
invariant under coordinate transformation.

\indent(v). Finally the function $\tilde{f}(r_H)$ is evaluated for
the solution of the equations of motion, i.e. all the fields:
$\{g_{ab}, \Phi_s, F^I_{ab}\}$ are on shell. For example, if the
near horizon geometry has the form
\begin{equation}
\label{nearhorizongeometry}
ds^2=v_1(-\rho^2d\tau^2+\frac{1}{\rho^2}d\rho^2)+v_2d\Omega_{n-2}^2\,
,
\end{equation}
and the equations of motion are satisfied, then we can express the
entropy in the form  (\ref{entropyfunction1}). There  $v_1$ and $
v_2$ should equal to $2/N''(r_H)$ and $\gamma(r_H)$.  $N$ ,
$\gamma$, and other fields, should satisfy the equations of motion.

One may worry about that the conserved charge form $\mathbf{Q}$ in
Eq.(\ref{conservedcharge}) is not complete: For example, we will
have an additional term $\mbox{{\boldmath $\epsilon$}}_{aba_1\cdots
a_{n-2}}\xi^a\nabla^{b}D(\phi)$ if the action has a dilaton coupling
term $D(\phi)R$. In general, the conserved charge form can be
written as $ \mathbf{Q}=\mathbf{Q}^{F}+\mathbf{Q}^g+\xi^a
\mathbf{W}_{a}+\mathbf{Y}+d\mathbf{Z} $, where $\mathbf{W}_a$,
$\mathbf{Y}$ and $\mathbf{Z}$ are smooth functions of fields and
their derivatives, and $\mathbf{Y}=\mathbf{Y}(\psi,
\mathcal{L}_{\xi}\psi)$ is linear for the field
variation~\cite{Iwald, Iwald1}. Obviously, $\mathbf{Y}$ and
$d\mathbf{Z}$ will not give contributions to the near horizon
integration ~(\ref{nhint}) if $\xi$ is a killing vector. It seems
that $\xi^a\mathbf{W}_a$ will give an additional contribution to
this integration. For the extremal case, this contribution will
vanish due to the smoothness of $\mathbf{W}_a$ and the vanishing
surface gravity. For example, the term corresponding to the dilaton
coupling mentioned above will vanish in the near horizon
integration. So the final form of the entropy~(\ref{entropynew})
will not change. For the non-extremal case, this term essentially
appear in the near horizon integration if we add the
$\xi^a\mathbf{W}_a$ into $\mathbf{Q}$. However, if necessary, we can
always change the Lagrangian $\mathbf{L}$ to be
$\mathbf{L}+d\mbox{{\boldmath $\mu$}}$ and put the conserved charge
form $\mathbf{Q}$ into the form of~(\ref{conservedcharge}) without
the $``\cdots"$ terms, where $\mbox{{\boldmath $\mu$}}$ is a
$(n-1)$-form. This change of Lagrangian will not affect the
equations of motion and the entropy of the black holes~\cite{Iwald,
Iwald1}. Then, the formulas (\ref{intnearhorzion}) and therefore
(\ref{nonextremal}) are still formally correct for the non-extremal
case after considering that ambiguity of the Lagrangian and
therefore $\tilde{f}(r_H)$. But this ambiguity has no contribution
to Eq. (\ref{entropynew}) which describes the entropy of the black
hole in the extremal case.

\section{Entropy function and attractor mechanism}

In this section  we show further that one can define an entropy
function with the help of the entropy definition of Wald. The
Noether current can always be written as
$\mathbf{J}[\xi]=d\mathbf{Q}[\xi]+\xi^a\mathbf{C}_a $ where
$\mathbf{C}_a$ corresponds to constraint. The constraint for the
$U(1)$ gauge fields  is (\ref{U1constrain}). If the equations of
motion for the $U(1)$ gauge fields hold, this constraint vanishes.
In this section, we will assume the equations of motion for the
$U(1)$ gauge fields are always satisfied, but not for the metric and
scalar fields. In other word, we will not consider the constraint
for the gauge fields. Assuming that the metric of the extremal black
holes has the form
$$
ds^2=-N(r)dt^2+\frac{1}{N(r)}dr^2+\gamma(r)d\Omega_{n-2}^2\, ,
$$
on the horizon $r=r_H$ of an extremal black hole, one has
$N(r_H)=0$, $N'(r_H)=0$, but $ N''(r_H) \ne 0$. Thus the near
horizon geometry will be fixed if $N''(r_H)$ and $\gamma(r_H)$ are
specified. This means the ``off-shell" of the near horizon geometry
corresponds to the arbitrariness of the parameter $N''(r_H)$ and
$\gamma(r_H)$. In the near horizon region ranged from $r_H$ to
$r_H+\Delta r$, we have
\begin{eqnarray}
&&\int_{r_{H}+\Delta
r}\mathbf{Q}[\xi]-\int_{r_{H}}\mathbf{Q}[\xi]+\int_{\mathcal{H}\times
\Delta r}\xi^a\mathbf{C}_a\nonumber \\
&&=\int_{\mathcal{H}\times \Delta
r}\mathbf{J}[\xi]=\int_{\mathcal{H}\times \Delta
r}\mathbf{\Theta}-\xi\cdot \mathbf{L}\, .
\end{eqnarray}
If $\xi$ is a Killing vector for the field configuration space for
our discussion (the solution space is a subset of this space), then
$\mathbf{\Theta}=0$, and we have
\begin{equation}
\int_{r_{H}+\Delta
r}\mathbf{Q}[\xi]-\int_{r_{H}}\mathbf{Q}[\xi]+\int_{\mathcal{H}\times
\Delta r}\xi^a\mathbf{C}_a=-\int_{\mathcal{H}\times \Delta
r}\xi\cdot \mathbf{L}\, .
\end{equation}
With this, we obtain
\begin{eqnarray}
\label{QgQgC} &&\int_{r_{H}+\Delta
r}\mathbf{Q}^g[\xi]-\int_{r_{H}}\mathbf{Q}^g[\xi]+\int_{\mathcal{H}\times
\Delta r}\xi^a\mathbf{C}_a\nonumber \\
&=&-\int_{r_{H}+\Delta
r}\mathbf{Q}^F[\xi]+\int_{r_{H}}\mathbf{Q}^F[\xi]-\int_{\mathcal{H}\times
\Delta r}\xi\cdot \mathbf{L}\, .
\end{eqnarray}
Define our  ``entropy function" as
\begin{equation}
\mathcal{E}=\lim_{\Delta r\rightarrow 0}\frac{4\pi}{N''(r_H)\Delta
r}\Bigg{\{}\int_{r_{H}+\Delta
r}\mathbf{Q}^g[\partial_t]-\int_{r_{H}}\mathbf{Q}^g[\partial_t]+\int_{\mathcal{H}\times
\Delta r}(\partial_t)^a\mathbf{C}_a\Bigg{\}}\, .
\end{equation}
If the equations of motion are satisfied, obviously, this
$\mathcal{E}$ will reduce to the entropy of extremal black holes
given in the previous section. Therefore this definition is
meaningful. Further, from Eq.~(\ref{QgQgC}), we have
\begin{equation}
\mathcal{E}=\lim_{\Delta r\rightarrow 0}\frac{4\pi}{N''(r_H)\Delta
r}\Bigg{\{}-\int_{r_{H}+\Delta
r}\mathbf{Q}^F[\partial_t]+\int_{r_{H}}\mathbf{Q}^F[\partial_t]-\int_{\mathcal{H}\times
\Delta r}\partial_t\cdot \mathbf{L}\Bigg{\}}\, .
\end{equation}
Recalling that the equations of motion for the $U(1)$ gauge fields
have been assumed to hold always, and following the calculations in
the previous section, we have
\begin{equation}
\mathcal{E}=\frac{4\pi}{N''(r_H)} \left(\tilde{e}_I
q_I-\tilde{f}(r_H)\right)\, .
\end{equation}
This expression looks the same as the one given  in the previous
section. However, a crucial difference from the one in the previous
section  is that here the fields need not be the solutions of the
equations of motion. To give the entropy of the extremal black
holes, we have to solve the equations of motion or extremize the
entropy function with respect to the undetermined values of fields
on the horizon.  It is easy to find that entropy function has the
form
\begin{equation}
\label{entropyfunction2} \mathcal{E}=\mathcal{E}(N'', \gamma, u_s,
\tilde{e}_I; p_i)=\frac{4\pi}{N''} \left(\tilde{e}_I
q_I-\tilde{f}_H(N'', \gamma, u_s, \tilde{e}_I; p_i)\right)\, ,
\end{equation}
where, for simplicity, we have denoted the $N''(r_H)$ and
$\gamma(r_H)$ by $N''$ and $\gamma$, respectively. The terms $u_s'$
will not appear because those kinetic terms of scalar fields in the
action always have a vanishing factor $N(r_H)=0$ on the horizon.
Similarly, $\gamma'(r_H),~\gamma''(r_H)$ will not appear because
that the components of the Riemann tensor which include these terms
have to contract with the vanished factors $N(r_H)$ or $N'(r_H)$.
Certainly, this point can be directly understood from the near
horizon geometry in Eq.~(\ref{nearhorizongeometry1}). So,
extremizing the entropy function becomes
\begin{equation}
\frac{\partial \mathcal{E}}{\partial N''}=0,\quad \frac{\partial
\mathcal{E}}{\partial \gamma}=0, \quad \frac{\partial
\mathcal{E}}{\partial u_s}=0\, .
\end{equation}
The electric charges are determined by
\begin{equation}
\frac{\partial \mathcal{E}}{\partial \tilde{e}_I}=0\quad \mathrm{or}
\quad q_I=\frac{\partial \tilde {f}(r_H)}{\partial \tilde{e}_I}\, .
\end{equation}
The entropy of the black hole can be obtained by solving these
algebraic equations, and substituting the solutions for $N''$,
$\gamma$, $u_s$ back into the entropy function.  If the values of
moduli fields on the horizon are determined by charges of black
holes, then the attractor mechanism is manifest. Then the entropy
has the form
\begin{equation}
S_{BH}=S_{BH}(q_I;
p_i)=\mathcal{E}|_{\mathrm{extremum}~\mathrm{piont}}\, ,
\end{equation}
a topological quantity which is fully determined by
charges~\cite{Sen2,Sen3}. These definitions will become more simple
if one chooses the coordinates $\{\tau, \rho,\cdots\}$ so that one
can define
\begin{equation}
v_1=\frac{2}{N''(r_H)},\quad v_2=\gamma(r_H)\, ,
\end{equation}
then, the entropy function can be written as
\begin{equation}
\mathcal{E}=\mathcal{E}( v_1, \,v_2, \,u_s, \, e_I; \,p_i)=2\pi
\left(e_Iq_I-f(v_1, \,v_2, \,u_s, \, e_I; \,p_i) \right)\, ,
\end{equation}
where $e_I$ are gauge fields on the horizon in this set of
coordinates, and $q_I=\frac{\partial f}{\partial e_I}$ are electric
charges which are not changed with the coordinate transformation.
So, in this set of coordinates, our entropy function form reduces to
the entropy function defined by A. Sen~\cite{Sen2,Sen3}.

\section{Conclusion and discussion}
In this paper, we have shown that the ``entropy function" method
proposed by A. Sen can be extracted from the general black hole
entropy definition of Wald~\cite{wald}.  For a spherically symmetry
extremal black hole as described by metric (\ref{generalmetric}), we
find that the entropy of the black hole can be put into a form
$$S_{BH}=\frac{4\pi }{N''(r_H)}\left(\tilde{e}_Iq_I-\tilde{f}(r_H)\right)$$
which is similar to the one given in Ref.~\cite{Sen2,Sen3}. To get
this entropy form, we have regarded the extremal black hole as the
extremal limit of an non-extremal black hole, i.e., we have required
(and only required) that the surface gravity approaches to zero. In
a special set of coordinates, i.e., $\{\tau,\rho\cdots\}$, this
entropy is exactly of the same form as the one given by A. Sen. We
have obtained a corresponding entropy function
(\ref{entropyfunction2}). After extremizing this entropy function
with respect to $N''$, $\gamma$ and other scalar fields, one gets
the entropy of the extremal black holes. Similarly, in the
coordinates $\{\tau,\rho\cdots \}$, our entropy function reduces to
the form of A. Sen. Note that in our procedure, we have neither used
the treatment of rescaling $AdS_2$ part of the near horizon geometry
of extremal black holes, nor especially employed the form of the
metric in the coordinates $\{\tau,\rho,\cdots\}$ as Eq.(\ref{eq1}).
In this procedure, it can be clearly seen why the electric charge
terms $e_Iq_I$ appear, but not the magnetic charges terms in the
entropy function.

Recently it was shown that for some near-extremal black holes with
BTZ black holes being a part of the near horizon geometry, that the
``entropy function" method works as well~\cite{CP3}. A similar
discussion for non-extremal $D3, M2$ and $M5$ branes has also been
given in~\cite{Garousi}. Therefore it is interesting to see whether
the procedure developed in this paper works or not for near-extremal
black holes. In this case, $N'(r_H)$ is an infinitesimal one instead
of vanishing. Eq. (\ref{nonextremal}) then gives
\begin{equation}
S_{BH}=2\pi B(r_H)=S_0\left(1+\frac{N'(r_H)}{
N''(r_H)}\frac{1}{r_*}\right)^{-1}\, ,
\end{equation}
where
\begin{equation}
S_0 = \frac{4\pi}{N''(r_H)}(\tilde{e}_Iq_I-\tilde{f}(r_H))\, ,
\end{equation}
and $r_*=B(r_H)/B'(r_H)$ approximately equals to
$``\frac{1}{n-2}\cdot \mathrm{radius~of~the~black~hole}"$ if the
higher derivative corrections in the effective action are small.
Thus, after considering that ambiguity in $\tilde{f}(r_H)$ becomes
very small and for large $r_*$ (sometimes, this corresponds to large
charges), the entropy function method gives us an approximate
entropy for near-extremal black holes, but the attractor mechanism
will be destroyed~\cite{DST}. In addition, it is also interesting to
discuss the extremal rotating black holes with the procedure
developed in this paper. Certainly, in this case, the Killing vector
which generates the horizon should be of the form
$\chi=\partial_t+\Omega_H
\partial_{\phi}$ instead of $\xi=\partial_t$.
A term including angular-momentum $J$ will naturally appear in the
associated entropy function~\cite{AGJST}. This issue is under
investigation.

\section*{Acknowledgements}
L.M.Cao thanks Hua Bai, Hui Li, Da-Wei Pang, Ding Ma, Yi Zhang and
Ya-Wen Sun for useful discussions and kind help. This work is
supported by grants from NSFC, China (No. 10325525 and No.
90403029), and a grant from the Chinese Academy of Sciences.

\end{document}